# Statistical mechanics of neocortical interactions:
# Training and testing canonical momenta indicators of EEG

Lester Ingber

*Lester Ingber Research, PO Box 06440, Wacker Dr PO - Sears Tower, Chicago, IL 60606-0440*

and

*DRW Investments LLC, Chicago Mercantile Exchange Center, 30 S Wacker Dr Ste 1516, Chicago, IL 60606*

ingber@ingber.com, ingber@alumni.caltech.edu

**Abstract**—A series of papers has developed a statistical mechanics of neocortical interactions (SMNI), deriving aggregate behavior of experimentally observed columns of neurons from statistical electrical-chemical properties of synaptic interactions. While not useful to yield insights at the single neuron level, SMNI has demonstrated its capability in describing large-scale properties of short-term memory and electroencephalographic (EEG) systematics. The necessity of including nonlinear and stochastic structures in this development has been stressed. Sets of EEG and evoked potential data were fit, collected to investigate genetic predispositions to alcoholism and to extract brain "signatures" of short-term memory. Adaptive Simulated Annealing (ASA), a global optimization algorithm, was used to perform maximum likelihood fits of Lagrangians defined by path integrals of multivariate conditional probabilities. Canonical momenta indicators (CMI) are thereby derived for individual's EEG data. The CMI give better signal recognition than the raw data, and can be used to advantage as correlates of behavioral states. These results give strong quantitative support for an accurate intuitive picture, portraying neocortical interactions as having common algebraic or physics mechanisms that scale across quite disparate spatial scales and functional or behavioral phenomena, i.e., describing interactions among neurons, columns of neurons, and regional masses of neurons. This paper adds to these previous investigations two important aspects, a description of how the CMI may be used in source localization, and calculations using previously ASA-fitted parameters in out-of-sample data.

**Keywords**—Electroencephalography, EEG, Simulated Annealing, Statistical Mechanics

## 1. INTRODUCTION

A model of statistical mechanics of neocortical interactions (SMNI) has been developed [1-21], describing large-scale neocortical activity on scales of mm to cm as measured by scalp EEG, with an audit trail back to minicolumnar interactions among neurons. This paper adds to these previous investigations two important aspects, a description of how the CMI may be used in source localization, and calculations using previously ASA-fitted parameters in out-of-sample data.

The underlying mathematical physics used to develop SMNI gives rise to a natural coordinate system faithful to nonlinear multivariate sets of potential data such as measured by multi-electrode EEG, canonical momenta indicators (CMI) [20,22,23]. Recent papers in finance [22,23] and in EEG systems [21] have demonstrated that CMI give enhanced signal resolutions over raw data.

The basic philosophy of SMNI is that good physical models of complex systems, often detailed by variables not directly measurable in many experimental paradigms, should offer superior descriptions of empirical data beyond that available from black-box statistical descriptions of such data. For example, good nonlinear models often offer sound approaches to relatively deeper understandings of these systems in terms of synergies of subsystems at finer spatial-temporal scales.

In this context, a generic mesoscopic neural network (MNN) has been developed for diffusion-type systems using a confluence of techniques drawn from the SMNI, modern methods of functional stochastic calculus defining nonlinear Lagrangians, adaptive simulated annealing (ASA) [24], and parallel-processing computation, to develop MNN [14,19]. MNN increases the resolution of SMNI to minicolumnar interactions within and between neocortical regions, a scale that overlaps with other studies of neural systems, e.g., artificial neural networks (ANN).

In order to interface the algebra presented by SMNI with experimental data, several codes have been developed. A key tool is adaptive simulated annealing (ASA), a global optimization C-language code [24-28]. Over the years, this code has evolved to a high degree of robustness across many disciplines. However, there are over 100 OPTIONS available for tuning this code; this is as expected for any single global optimization code applicable to many classes of nonlinear systems, systems which typically are non-typical.

Section 2 gives the background used to develop SMNI and ASA for the present study. Appendix A gives more detail on ASA relevant to this paper. Section 3 gives the mathematical development required for this study. Section 4 describes the procedures used. Section 5 discusses source localization in the context of global and local EEG theories. Section 6 presents conclusions.

## 2. BACKGROUND

### 2.1. EEG

The SMNI approach develops mesoscopic scales of neuronal interactions at columnar levels of hundreds of neurons from the statistical mechanics of relatively microscopic interactions at neuronal and synaptic scales, poised to study relatively macroscopic dynamics at regional scales as measured by scalp electroencephalography (EEG). Relevant experimental confirmation is discussed in the SMNI papers at the mesoscopic scales as well as at macroscopic scales of scalp EEG. The derived firings of columnar activity, considered as order parameters of the mesoscopic system, develop multiple attractors, which illuminate attractors that may be present in the macroscopic regional dynamics of neocortex. SMNI proposes that models to be fitted to data include models of activity under each electrode, e.g., due to short-ranged neuronal fibers, as well as models of activity across electrodes, e.g., due to long-ranged fibers. These influences can be disentangled by SMNI fits.

The SMNI approach is complementary to other methods of studying nonlinear neocortical dynamics at macroscopic scales. For example, EEG and magnetoencephalography (MEG) data have been expanded in a series of spatial principal components, a Karhunen-Loeve expansion. The coefficients in such expansions are identified as order parameters that characterize phase changes in cognitive studies [29,30] and epileptic seizures [31,32]. However, the SMNI CMI may be considered in a similar context, as providing a natural coordinate system that can be sensitive to experimental data, without assuming averages over stochastic parts of the system that may contain important information.

Theoretical studies of the neocortical medium have involved local circuits with postsynaptic potential delays [33-36], global studies in which finite velocity of action potential and periodic boundary conditions are important [37-40], and nonlinear nonequilibrium SMNI. The local and the global theories combine naturally to form a single theory in which control parameters effect changes between more local and more global dynamic behavior [40,41], in a manner somewhat analogous to localized and extended wave-function states in disordered solids.

Plausible connections between the multiple-scale statistical theory and the more phenomenological global theory have been proposed [12]. Experimental studies of neocortical dynamics with EEG include maps of magnitude distribution over the scalp [38,42], standard Fourier analyses of EEG time series [38], and estimates of correlation dimension [43,44]. Other studies have emphasized that many EEG states are accurately described by a few coherent spatial modes exhibiting complex temporal behavior [29-32,38,40]. These modes are the order parameters at macroscopic scales that underpin the phase changes associated with changes of physiological state.

For extracranial EEG, it is clear that spatial resolution, i.e., the ability to distinguish between two dipole sources as their distance decreases, is different from dipole localization, i.e., the ability to locate a single dipole [40]. The development of methods to improve the spatial resolution of EEG has made it more practical to study spatial structure. For example, high resolution methods provide apparent spatial resolution in the 2-3 cm range [45]. Dipole resolution may be as good as several mm [46]. Some algorithms calculate the (generally non-unique) inverse-problem of determining cortical sources that are weighted/filtered by volume conductivities of concentric spheres encompassing the brain, cerebrospinal fluid, skull, and scalp. A straightforward approach is to calculate the surface Laplacian from spline fits to the scalp potential distribution, yielding estimates similar to those obtained using concentric spheres models of the head [45]. Other measuring techniques, e.g., MEG, can provide complementary information. These methods have their strengths and weaknesses at various spatial-temporal frequencies.

These source localization methods typically do not include in their models the synergistic contributions from short-ranged columnar firings of mm spatial extent and from long-ranged fibers spanning cm spatial extent. The CMI study presented here models these synergistic short-ranged and long-ranged interactions.

### 2.2. Short-Term Memory (STM)

The development of SMNI in the context of short-term memory (STM) tasks leads naturally to the identification of measured electric scalp potentials as arising from excitatory and inhibitory short-ranged and excitatory long-ranged fibers as they contribute to minicolumnar interactions [12,13]. Therefore, the SMNI CMI are most appropriately calculated in the context of STM experimental paradigms. It has been been demonstrated that EEG data from such paradigms can be fit using only physical synaptic and neuronal parameters that lie within experimentally observed ranges [13,20].

The SMNI calculations are of minicolumnar interactions among hundreds of neurons, within a macrocolumnar extent of hundreds of thousands of neurons. Such interactions take place on time scales of several $\tau$, where $\tau$ is on the order of 10 msec (of the order of time constants of cortical pyramidal cells). This also is the observed time scale of the dynamics of STM. SMNI hypothesizes that columnar interactions within and/or between regions containing many millions of neurons are responsible for these phenomena at time scales of several seconds. That is, the nonlinear evolution at finer temporal scales gives a base of support for the phenomena observed at the coarser temporal scales, e.g., by establishing mesoscopic attractors at many macrocolumnar spatial locations to process patterns in larger regions.

SMNI has presented a model of STM, to the extent it offers stochastic bounds for this phenomena during focused selective attention [4,6,15,47-49], transpiring on the order of tenths of a second to seconds, limited to the retention of $7 \pm 2$ items [50]. These constraints exist even for apparently exceptional memory performers who, while they may be capable of more efficient encoding and retrieval of STM, and while they may be more efficient in "chunking" larger patterns of information into single items, nevertheless are limited to a STM capacity of $7 \pm 2$ items [51]. Mechanisms for various STM phenomena have been proposed across many spatial scales [52]. This "rule" is verified for acoustical STM, as well as for visual or semantic STM, which typically require longer times for rehearsal in an hypothesized articulatory loop of individual items, with a capacity that appears to be limited to $4 \pm 2$ [53]. SMNI has detailed these constraints in models of auditory and visual cortex [4,6,15,16].

Another interesting phenomenon of STM capacity explained by SMNI is the primacy versus recency effect in STM serial processing [6], wherein first-learned items are recalled most error-free, with last-learned items still more error-free than those in the middle [54]. The basic assumption being made is that a pattern of neuronal firing that persists for many $\tau$ cycles is a candidate to store the "memory" of activity that gave rise to this pattern. If several firing patterns can simultaneously exist, then there is the capability of storing several memories. The short-time probability distribution derived for the neocortex is the primary tool to seek such firing patterns. The deepest minima are more likely accessed than the others of this probability distribution, and these valleys are sharper than the others. I.e., they are more readily accessed and sustain their patterns against fluctuations more accurately than the others. The more recent memories or newer patterns may be presumed to be those having synaptic parameters more recently tuned and/or more actively rehearsed.

It has been noted that experimental data on velocities of propagation of long-ranged fibers [38,40] and derived velocities of propagation of information across local minicolumnar interactions [2] yield comparable times scales of interactions across minicolumns of tenths of a second. Therefore, such phenomena as STM likely are inextricably dependent on interactions at local and global scales.

### 2.2.1. SMNI & ADP

A proposal has been advanced that STM is processed by information coded in approximately 40-Hz (approximately 2.5 foldings of $\tau$) bursts per stored memory, permitting up to seven such memories to be processed serially within single waves of lower frequencies on the order of 5 to 12 Hz [55]. To account for the observed duration of STM, they propose that observed after-depolarization (ADP) at synaptic sites, affected by the action of relatively long-time acting neuromodulators, e.g., acetylcholine and serotonin, acts to regularly "refresh" the stored memories in subsequent oscillatory cycles. A recent study of the action of neuromodulators in the neocortex supports the premise of their effects on broad spatial and temporal scales [56], but the ADP model is much more specific in its proposed spatial and temporal influences.

SMNI does not detail any specific synaptic or neuronal mechanisms that might refresh these most-likely states to reinforce multiple short-term memories [18]. However, the calculated evolution of states is consistent with the observation that an oscillatory subcycle of 40 Hz may be the bare minimal threshold of self-sustaining minicolumnar firings before they begin to degrade [16].

The mechanism of ADP details a specific synaptic mechanism that, when coupled with additional proposals of neuronal oscillatory cycles of 5–12 Hz and oscillatory subcycles of 40 Hz, can sustain these memories for longer durations on the order of seconds. By itself, ADP does not provide a constraint such as the 7±2 rule. The ADP approach does not address the observed random access phenomena of these memories, the 4±2 rule, the primacy versus recency rule, or the influence of STM in observed EEG patterns.

SMNI and ADP models are complementary to the understanding of STM. MNN can be used to overlap the spatial scales studied by the SMNI with the finer spatial scales typically studied by other relatively more microscopic neural networks. At this scale, such models as ADP are candidates for providing an extended duration of firing patterns within the microscopic networks.

### 2.2.2. PATHINT

A path-integral C-language code, PATHINT, calculates the long-time probability distribution from the Lagrangian, e.g., as fit by the ASA code. A robust and accurate histogram-based (non-Monte Carlo) path-integral algorithm to calculate the long-time probability distribution had been developed to handle nonlinear Lagrangians [57-59], which was extended to two-dimensional problems [60]. PATHINT was developed for use in arbitrary dimensions, with additional code to handle general Neumann and Dirichlet conditions, as well as the possibility of including time-dependent potentials, drifts, and diffusions. The results of using PATHINT to determine the evolution of the attractors of STM give overall results consistent with previous calculations [15,16].

### 2.3. ASA

In order to maintain some audit trail from large-scale regional activity back to mesoscopic columnar dynamics, desirable for both academic interest as well as practical signal enhancement, as few approximations as possible are made by SMNI in developing synaptic interactions up to the level of regional activity as measured by scalp EEG. This presents a formidable multivariate nonlinear nonequilibrium distribution as a model of EEG dynamics, a concept considered to be quite tentative by research panels as late as 1990, until it was demonstrated how fits to EEG data could be implemented [13].

In order to fit such distributions to real data, ASA has been developed, a global optimization technique, a superior variant of simulated annealing [25]. This was tested using EEG data in 1991 [13], using an early and not as flexible version of ASA, very fast reannealing (VFSR) [25]. Here, this is tested on more refined EEG using more sensitive CMI to portray results of the fits [20].

ASA [24] fits short-time probability distributions to observed data, using a maximum likelihood technique on the Lagrangian. This algorithm has been developed to fit observed data to a theoretical cost function over a $D$-dimensional parameter space [25], adapting for varying sensitivities of parameters during the fit.

Appendix A contains details of ASA relevant to its use in this paper.

## 2.4. Complementary Research

### 2.4.1. Chaos

Given the context of studies in complex nonlinear systems [61], the question can be asked: What if EEG has chaotic mechanisms that overshadow the above stochastic considerations? The real issue is whether the scatter in data can be distinguished between being due to noise or chaos [62]. In this regard, several studies have been proposed with regard to comparing chaos to simple filtered (colored) noise [61,63]. Since the existence of multiplicative noise in neocortical interactions has been derived, then the previous references must be generalized, and further investigation is required to decide whether EEG scatter can be distinguished from multiplicative noise.

A recent study with realistic EEG wave equations strongly suggests that if chaos exists in a deterministic limit, it does not survive in macroscopic stochastic neocortex [64]. I.e., it is important to include stochastic aspects, as arise from the statistics of synaptic and columnar interactions, in any realistic description of macroscopic neocortex.

### 2.4.2. Other Systems

Experience using ASA on such multivariate nonlinear stochastic systems has been gained by similar applications of the approach used for SMNI.

From 1986-1989, these methods of mathematical physics were utilized by a team of scientists and officers to develop mathematical comparisons of Janus computer combat simulations with exercise data from the National Training Center (NTC), developing a testable theory of combat successfully baselined to empirical data [60,65-69].

This methodology has been applied to financial markets [22,23,70-72], developing specific trading rules for S&P 500 to demonstrate the robustness of these mathematical and numerical algorithms.

## 3. MATHEMATICAL DEVELOPMENT

Fitting a multivariate nonlinear stochastic model to data is a necessary, but not sufficient procedure in developing new diagnostic software. Even an accurate model fit well to real data may not be immediately useful to clinicians and experimental researchers. To fill this void, the powerful intuitive basis of the mathematical physics used to develop SMNI has been utilized to describe the model in terms of rigorous CMI that provide an immediate intuitive portrait of the EEG data, faithfully describing the neocortical system being measured. The CMI give an enhanced signal over the raw data, and give some insights into the underlying columnar interactions.

### 3.1. CMI, Information, Energy

In the first SMNI papers, it was noted that this approach permitted the calculation of a true nonlinear nonequilibrium "information" entity at columnar scales. With reference to a steady state $\bar{P}(\tilde{M})$ for a short-time Gaussian-Markovian conditional probability distribution $P$ of variables $\tilde{M}$, when it exists, an analytic definition of the information gain $\hat{\Upsilon}$ in state $\tilde{P}(\tilde{M})$ over the entire neocortical volume is defined by [73,74]

$$\hat{\Upsilon}[\tilde{P}] = \int \cdots \int D\tilde{M} \; \tilde{P} \ln(\tilde{P}/\bar{P}) \; , \quad D M = (2\pi \hat{g}_0^2 \Delta t)^{-1/2} \prod_{s=1}^{u} (2\pi \hat{g}_s^2 \Delta t)^{-1/2} dM_s \; , \tag{1}$$

where a path integral is defined such that all intermediate-time values of $\tilde{M}$ appearing in the folded short-time distributions $\tilde{P}$ are integrated over. This is quite general for any system that can be described as Gaussian-Markovian [75], even if only in the short-time limit, e.g., the SMNI theory.

As time evolves, the distribution likely no longer behaves in a Gaussian manner, and the apparent simplicity of the short-time distribution must be supplanted by numerical calculations. The Feynman Lagrangian is written in the midpoint discretization, for a specific macrocolumn corresponding to

$$M(\bar{t}_s) = \frac{1}{2}[M(t_{s+1}) + M(t_s)] \; . \tag{2}$$

This discretization defines a covariant Lagrangian $L_F$ that possesses a variational principle for arbitrary noise, and that explicitly portrays the underlying Riemannian geometry induced by the metric tensor $g_{GG'}$, calculated to be the inverse of the covariance matrix $g^{GG'}$. Using the Einstein summation convention,

$$P = \int \cdots \int DM \exp\left(-\sum_{s=0}^{u} \Delta t L_{Fs}\right),$$

$$DM = g_{0+}^{1/2}(2\pi\Delta t)^{-\Theta/2} \prod_{s=1}^{u} g_{s+}^{1/2} \prod_{G=1}^{\Theta} (2\pi\Delta t)^{-1/2} dM_s^G \; ,$$

$$\int dM_s^G \to \sum_{\iota=1}^{N^G} \Delta M_{\iota s}^G, \, M_0^G = M_{t_0}^G, \, M_{u+1}^G = M_t^G,$$

$$L_F = \frac{1}{2}(dM^G/dt - h^G)g_{GG'}(dM^{G'}/dt - h^{G'}) + \frac{1}{2}h^G{}_{;G} + R/6 - V,$$

$$(\cdots)_{,G} = \frac{\partial(\cdots)}{\partial M^G},$$

$$h^G = g^G - \frac{1}{2}g^{-1/2}(g^{1/2}g^{GG'})_{,G'},$$

$$g_{GG'} = (g^{GG'})^{-1},$$

$$g_s[M^G(\bar{t}_s), \bar{t}_s] = \det(g_{GG'})_s, \, g_{s_+} = g_s[M_{s+1}^G, \bar{t}_s],$$

$$h^G{}_{;G} = h^G{}_{,G} + \Gamma^F_{GF}h^G = g^{-1/2}(g^{1/2}h^G)_{,G},$$

$$\Gamma^F_{JK} \equiv g^{LF}[JK, L] = g^{LF}(g_{JL,K} + g_{KL,J} - g_{JK,L}),$$

$$R = g^{JL}R_{JL} = g^{JL}g^{JK}R_{FJKL},$$

$$R_{FJKL} = \frac{1}{2}(g_{FK,JL} - g_{JK,FL} - g_{FL,JK} + g_{JL,FK}) + g_{MN}(\Gamma^M_{FK}\Gamma^N_{JL} - \Gamma^M_{FL}\Gamma^N_{JK}), \quad (3)$$

where $R$ is the Riemannian curvature, and the discretization is explicitly denoted in the mesh of $M_{\iota s}^G$ by $\iota$. If $M$ is a field, e.g., also dependent on a spatial variable $x$ discretized by $\nu$, then the variables $M_s^G$ is increased to $M_s^{G\nu}$, e.g., as prescribed for the macroscopic neocortex. The term $R/6$ in $L_F$ includes a contribution of $R/12$ from the WKB approximation to the same order of $(\Delta t)^{3/2}$ [76].

A prepoint discretization for the same probability distribution $P$ gives a much simpler algebraic form,

$$M(\bar{t}_s) = M(t_s),$$

$$L = \frac{1}{2}(dM^G/dt - g^G)g_{GG'}(dM^{G'}/dt - g^{G'}) - V, \quad (4)$$

but the Lagrangian $L$ so specified does not satisfy a variational principle useful for moderate to large noise; its associated variational principle only provides information useful in the weak-noise limit [77]. The neocortex presents a system of moderate noise. Still, this prepoint-discretized form has been quite useful in all systems examined thus far, simply requiring a somewhat finer numerical mesh. Note that although integrations are indicated over a huge number of independent variables, i.e., as denoted by $dM_s^{Gv}$, the physical interpretation afforded by statistical mechanics makes these systems mathematically and physically manageable.

It must be emphasized that the output need not be confined to complex algebraic forms or tables of numbers. Because $\underline{L}_F$ possesses a variational principle, sets of contour graphs, at different long-time epochs of the path-integral of $P$, integrated over all its variables at all intermediate times, give a visually intuitive and accurate decision aid to view the dynamic evolution of the scenario. For example, as given in Table 1, this Lagrangian approach permits a quantitative assessment of concepts usually only loosely defined [69,78]. In this study, the above canonical momenta are referred to canonical momenta indicators (CMI).

---

Table 1

---

In a prepoint discretization, where the Riemannian geometry is not explicit (but calculated in the first SMNI papers), the distributions of neuronal activities $p_{\sigma_i}$ is developed into distributions for activity under an electrode site $P$ in terms of a Lagrangian $\underline{L}$ and threshold functions $F^G$,

$$P = \prod_G P^G[M^G(r; t+\tau)|M^{\bar{G}}(r'; t)] = \sum_{\sigma_j} \delta\left(\sum_{jE} \sigma_j - M^E(r; t+\tau)\right) \delta\left(\sum_{jI} \sigma_j - M^I(r; t+\tau)\right) \prod_j^N p_{\sigma_j}$$

$$\approx \prod_G (2\pi\tau g^{GG})^{-1/2} \exp(-N\tau \underline{L}^G) = (2\pi\tau)^{-1/2} g^{1/2} \exp(-N\tau \underline{L}) ,$$

$$\underline{L} = \underline{T} - \underline{V} \ , \ \underline{T} = (2N)^{-1}(\dot{M}^G - g^G) g_{GG'}(\dot{M}^{G'} - g^{G'}) ,$$

$$g^G = -\tau^{-1}(M^G + N^G \tanh F^G) \ , \ g^{GG'} = (g_{GG'})^{-1} = \delta_G^{G'} \tau^{-1} N^G \text{sech}^2 F^G \ , \ g = \det(g_{GG'}) ,$$

$$F^G = \frac{V^G - v_{G'}^{|G|} T_{G'}^{|G|}}{(\pi[(v_{G'}^{|G|})^2 + (\phi_{G'}^{|G|})^2] T_{G'}^{|G|})^{1/2}} ,$$

$$T_{G'}^{|G|} = a_{G'}^{|G|} N^{G'} + \frac{1}{2} A_{G'}^{|G|} M^{G'} + a_{G'}^{\dagger|G|} N^{\dagger G'} + \frac{1}{2} A_{G'}^{\dagger|G|} M^{\dagger G'} + a_{G'}^{\ddagger|G|} N^{\ddagger G'} + \frac{1}{2} A_{G'}^{\ddagger|G|} M^{\ddagger G'} ,$$

$$a_{G'}^{\dagger G} = \frac{1}{2} A_{G'}^{\dagger G} + B_{G'}^{\dagger G} \ , \ A_E^{\ddagger I} = A_I^{\ddagger E} = A_I^{\ddagger I} = B_E^{\ddagger I} = B_I^{\ddagger E} = B_I^{\ddagger I} = 0 \ , \ a_E^{\ddagger E} = \frac{1}{2} A_E^{\ddagger E} + B_E^{\ddagger E} , \quad (5)$$

where no sum is taken over repeated $|G|$, $A^G_{G'}$ and $B^G_{G'}$ are macrocolumnar-averaged interneuronal synaptic efficacies, $v^G_{G'}$ and $\phi^G_{G'}$ are averaged means and variances of contributions to neuronal electric polarizations, $N^G$ are the numbers of excitatory and inhibitory neurons per minicolumn, and the variables associated with $M^G$, $M^{\dagger G}$ and $M^{\ddagger G}$ relate to multiple scales of activities from minicolumns, between minicolumns within regions, and across regions, resp. The nearest-neighbor interactions $\underline{V}$ can be modeled in greater detail by a stochastic mesoscopic neural network [14]. The SMNI papers give more detail on this derivation.

In terms of the above variables, an energy or Hamiltonian density $H$ can be defined,

$$\underline{H} = \underline{T} + \underline{V} , \quad (6)$$

in terms of the $M^G$ and $\Pi^G$ variables, and the path integral is now defined over all the $\underline{D}M^G$ as well as over the $\underline{D}\Pi^G$ variables.

### 3.2. Regional Propagator

Relevant to the issue of source localization is how SMNI describes individual head shapes. The algebra required to treat these issues was developed in the earliest SMNI papers [2,3].

Define the $\Lambda$-dimensional spatial vector $\tilde{M}_s$ at time $t_s$,

$$\tilde{M}_s = \{ M^\nu_s = M_s(r^\nu); \nu = 1, \cdots, \Lambda \} ,$$

$$M^\nu_s = \{ M^{G\nu}_s; G = E, I \} . \quad (7)$$

For macroscopic space-time considerations, mesoscopic $\rho$ (spatial extent of a minicolumn) and $\tau$ scales are measured by $dr$ and $dt$. In the continuum limits of $r$ and $t$,

$$M^{G\nu}_s \to M^G(r,t), \dot{M}^{G\nu}_s \to dM^G/dt,$$

$$(M^{G,\nu+1} - M^{G\nu})/(r^{\nu+1} - r^\nu) \to \nabla_r M^G . \quad (8)$$

The previous development of mesocolumnar interactions via nearest-neighbor derivative couplings permits the regional short-time propagator $\tilde{P}$ to be developed in terms of the Lagrangian $\underline{L}$ [79]:

$$\tilde{P}(\tilde{M}) = (2\pi\theta)^{-\Lambda/2} \int d\tilde{M} \, g^{\Lambda/2} \exp[-N\tilde{S}(\tilde{M})]\tilde{P}(\tilde{M}),$$

$$\tilde{S} = \min \int_t^{t+\theta} dt' L[\dot{M}(t'), M(t')],$$

$$L = \Lambda \Omega^{-1} \int d^2 r \, \underline{L}, \tag{9}$$

where $\Omega$ is the area of the region considered, and

$$\Lambda \Omega^{-1} \int d^2 r = \Lambda \Omega^{-1} \int dx\, dy = \lim_{\substack{\rho \to 0 \\ \Lambda \to \infty}} \sum_{\nu=1}^{\Lambda}. \tag{10}$$

With $\tilde{P}$ properly defined by this space-time mesh, a path-integral formulation for the regional long-time propagator at time $t = (u+1)\theta + t_0$ is developed:

$$\tilde{P}[\tilde{M}(t)]\, d\tilde{M}(t) = \int \cdots \int \underline{D}\tilde{M} \exp(-N \int_{t_0}^t dt' L),$$

$$\tilde{P}[\tilde{M}(t_0)] = \delta(\tilde{M} - \tilde{M}_0),$$

$$\underline{D}\tilde{M} = \prod_{s=1}^{u+1} \prod_{\nu=1}^{\Lambda} \prod_G^{E,I} (2\pi\theta)^{-\frac{1}{2}} (g_s^\nu)^{1/4} dM_s^{G\nu}. \tag{11}$$

Note that, even for $N\tau L \approx 1$, $N \int_{t_0}^t dt' L$ is very large for macroscopically large time $(t - t_0)$ and macroscopic size $\Lambda$, demonstrating how extrema of $L$ define peaked maximum probability states. This derivation can be viewed as containing the dynamics of macroscopic causal irreversibility, whereby $\tilde{P}$ is an unstable fixed point about which deviations from the extremum are greatly amplified [80].

Thus, any brain surface is mapped according to its $\Lambda$ minicolumns, wherein its geometry is explicitly developed according the the discretized path integral according to $d\tilde{M}$ over the minicolumnar-discretized surface mapped by $d^2 r$.

### 3.3. Nonlinear String Model

A mechanical-analog, the string model, is derived explicitly for neocortical interactions using SMNI [12]. In addition to providing overlap with current EEG paradigms, this defines a probability distribution of firing activity, which can be used to further investigate the existence of other nonlinear phenomena, e.g., bifurcations or chaotic behavior, in brain states.

Previous SMNI studies have detailed that maximal numbers of attractors lie within the physical firing space of $M^G$, consistent with experimentally observed capacities of auditory and visual STM, when a "centering" mechanism is enforced by shifting background conductivities of synaptic interactions, consistent with experimental observations under conditions of selective attention [4,6,15,16,81]. This leads to an effect of having all attractors of the short-time distribution lie along a diagonal line in $M^G$ space, effectively defining a narrow parabolic trough containing these most likely firing states. This essentially collapses the 2 dimensional $M^G$ space down to a 1 dimensional space of most importance.

Thus, the predominant physics of short-term memory and of (short-fiber contribution to) EEG phenomena takes place in a narrow "parabolic trough" in $M^G$ space, roughly along a diagonal line [4]. The object of interest within a short refractory time, $\tau$, approximately 5 to 10 msec, is the Lagrangian $\underline{L}$ for a mesocolumn, detailed above. $\tau \underline{L}$ can vary by as much as a factor of $10^5$ from the highest peak to the lowest valley in $M^G$ space. Therefore, it is reasonable to assume that a single independent firing variable might offer a crude description of this physics. Furthermore, the scalp potential $\Phi$ can be considered to be a function of this firing variable. (Here, "potential" refers to the electric potential, not the potential term in the Lagrangian above.) In an abbreviated notation subscripting the time-dependence,

$$\Phi_t - \ll \Phi \gg = \Phi(M_t^E, M_t^I) \approx a(M_t^E - \ll M^E \gg) + b(M_t^I - \ll M^I \gg) , \tag{12}$$

where $a$ and $b$ are constants, and $\ll \Phi \gg$ and $\ll M^G \gg$ represent typical minima in the trough. In the context of fitting data to the dynamic variables, there are three effective constants, $\{a, b, \phi\}$,

$$\Phi_t - \phi = aM_t^E + bM_t^I . \tag{13}$$

Accordingly, there is assumed to be a linear relationship (about minima to be fit to data) between the $M^G$ firing states and the measured scalp potential $\Phi_\nu$, at a given electrode site $\nu$ representing a macroscopic region of neuronal activity:

$$\Phi_\nu - \phi = aM^E + bM^I , \tag{14}$$

where $\{\phi, a, b\}$ are constants determined for each electrode site. In the prepoint discretization, the postpoint $M^G(t + \Delta t)$ moments are given by

$$m \equiv < \Phi_\nu - \phi > = a < M^E > + b < M^I > = ag^E + bg^I ,$$

$$\sigma^2 \equiv < (\Phi_\nu - \phi)^2 > - < \Phi_\nu - \phi >^2 = a^2 g^{EE} + b^2 g^{II} , \tag{15}$$

where the $M^G$-space drifts $g^G$, and diffusions $g^{GG'}$, are given above. Note that the macroscopic drifts and diffusions of the $\Phi$'s are simply linearly related to the mesoscopic drifts and diffusions of the $M^G$'s. For the prepoint $M^G(t)$ firings, the same linear relationship in terms of $\{\phi, a, b\}$ is assumed.

The mesoscopic probability distributions, $P$, are scaled and aggregated over this columnar firing space to obtain the macroscopic probability distribution over the scalp-potential space:

$$P_\Phi[\Phi] = \int dM^E dM^I P[M^E, M^I] \delta[\Phi - \Phi'(M^E, M^I)] . \tag{16}$$

The parabolic trough described above justifies a form

$$P_\Phi = (2\pi\sigma^2)^{-1/2} \exp(-\frac{\Delta t}{2\sigma^2} \int dx\, L_\Phi) ,$$

$$L_\Phi = \frac{\alpha}{2} |\partial\Phi/\partial t|^2 + \frac{\beta}{2} |\partial\Phi/\partial x|^2 + \frac{\gamma}{2} |\Phi|^2 + F(\Phi) , \tag{17}$$

where $F(\Phi)$ contains nonlinearities away from the trough, $\sigma^2$ is on the order of $1/N$ given the derivation of $\underline{L}$ above, and the integral over $x$ is taken over the spatial region of interest. In general, there also will be terms linear in $\partial\Phi/\partial t$ and in $\partial\Phi/\partial x$.

Previous calculations of EEG phenomena [5], show that the short-fiber contribution to the $\alpha$ frequency and the movement of attention across the visual field are consistent with the assumption that the EEG physics is derived from an average over the fluctuations of the system, e.g., represented by $\sigma$ in the above equation. I.e., this is described by the Euler-Lagrange equations derived from the variational principle possessed by $L_\Phi$ (essentially the counterpart to force equals mass times acceleration), more properly by the "midpoint-discretized" Feynman $L_\Phi$, with its Riemannian terms [2,3,11], Hence, the variational principle applies,

$$0 = \frac{\partial}{\partial t}\frac{\partial L_\Phi}{\partial(\partial\Phi/\partial t)} + \frac{\partial}{\partial x}\frac{\partial L_\Phi}{\partial(\partial\Phi/\partial x)} - \frac{\partial L_\Phi}{\partial \Phi} . \tag{18}$$

The result is

$$\alpha \frac{\partial^2 \Phi}{\partial t^2} + \beta \frac{\partial^2 \Phi}{\partial x^2} + \gamma\Phi - \frac{\partial F}{\partial \Phi} = 0 . \tag{19}$$

If there exist regions in neocortical parameter space such that $\beta/\alpha = -c^2$, $\gamma/\alpha = \omega_0^2$,

$$\frac{1}{\alpha}\frac{\partial F}{\partial \Phi} = -\Phi f(\Phi) , \tag{20}$$

and $x$ is taken to be one-dimensional, then the nonlinear string is recovered. Terms linear in $\partial\Phi/\partial t$ and in $\partial\Phi/\partial x$ in $L_\Phi$ can make other contributions, e.g., giving rise to damping terms.

The path-integral formulation has a utility beyond its deterministic Euler-Lagrange limit. This has been utilized to explicitly examine the long-time evolution of systems, to compare models to long-time correlations in data [60,68]. This use is being extended to other systems, in finance [71,72] and in EEG modeling as described here.

For the prepoint $M^E(t)$ firings, advantage is taken of the parabolic trough derived for the STM Lagrangian, and

$$M^I(t) = cM^E(t) , \qquad (21)$$

where the slope $c$ is set to the close approximate value determined by a detailed calculation of the centering mechanism [15],

$$A^E_E M^E - A^E_I M^I \approx 0 . \qquad (22)$$

This permits a complete transformation from $M^G$ variables to $\Phi$ variables.

Similarly, as appearing in the modified threshold factor $F^G$, each regional influence from electrode site $\mu$ acting at electrode site $\nu$, given by afferent firings $M^{\ddagger E}$, is taken as

$$M^{\ddagger E}_{\mu\to\nu} = d_\nu M^E_\mu(t - T_{\mu\to\nu}) , \qquad (23)$$

where $d_\nu$ are constants to be fitted at each electrode site, and $T_{\mu\to\nu}$ are the delay times estimated above for inter-electrode signal propagation, based on anatomical knowledge of the neocortex and of velocities of propagation of action potentials of long-ranged fibers, typically on the order of one to several multiples of $\tau = 5$ msec. Some terms in which $d$ directly affects the shifts of synaptic parameters $B^G_{G'}$ when calculating the centering mechanism also contain long-ranged efficacies (inverse conductivities) $B^{*E}_{E'}$. Therefore, the latter were kept fixed with the other electrical-chemical synaptic parameters during these fits. Future fits will experiment taking the $T$'s as parameters.

This defines the conditional probability distribution for the measured scalp potential $\Phi_\nu$,

$$P_\nu[\Phi_\nu(t+\Delta t)|\Phi_\nu(t)] = \frac{1}{(2\pi\sigma^2\Delta t)^{1/2}} \exp(-L_\nu\Delta t) ,$$

$$L_\nu = \frac{1}{2\sigma^2}(\dot\Phi_\nu - m)^2 . \qquad (24)$$

The probability distribution for all electrodes is taken to be the product of all these distributions:

$$P = \prod_\nu P_\nu \ , \ L = \sum_\nu L_\nu \ . \tag{25}$$

Note that the belief in the dipole or nonlinear-string model is being invoked. The model SMNI, derived for $P[M^G(t+\Delta t)|M^{\bar{G}}(t)]$, is for a macrocolumnar-averaged minicolumn; hence it is expected to be a reasonable approximation to represent a macrocolumn, scaled to its contribution to $\Phi_\nu$. Hence, $L$ is used to represent this macroscopic regional Lagrangian, scaled from its mesoscopic mesocolumnar counterpart $\underline{L}$. However, the above expression for $P_\nu$ uses the dipole assumption to also use this expression to represent several to many macrocolumns present in a region under an electrode: A macrocolumn has a spatial extent of about a mm. A scalp electrode has been shown, under extremely favorable circumstances, to have a resolution as small as several mm, directly competing with the best spatial resolution attributed to MEG [46]. It is often argued that typically several macrocolumns firing coherently account for the electric potentials measured by one scalp electrode [82]. Then, this model is being tested to see if the potential will scale to a representative macrocolumn. The results presented here seem to confirm that this approximation is in fact quite reasonable.

Future projects will develop SMNI to include higher resolution minicolumnar-minicolumnar dynamics using stochastic MNN, described above [14].

### 3.4. CMI Sensitivity

In the SMNI approach, "information" is a concept well defined in terms of the probability eigenfunctions of electrical-chemical activity of this Lagrangian. The path-integral formulation presents an accurate intuitive picture of an initial probability distribution of patterns of firings being filtered by the (exponential of the) Lagrangian, resulting in a final probability distribution of patterns of firing.

The utility of a measure of information has been noted by other investigators. For example, there have been attempts to use information as an index of EEG activity [83,84]. These attempts have focused on the concept of "mutual information" to find correlations of EEG activity under different electrodes. Other investigators have looked at simulation models of neurons to extract information as a measure of complexity of information processing [85]. Some other investigators have examined the utility of the energy density as a viable measure of information processing STM paradigms [86].

The SMNI approach at the outset recognizes that, for most brain states of late latency, at least a subset of regions being measured by several electrodes is indeed to be considered as one system, and their interactions are to be explicated by mathematical or physical modeling of the underlying neuronal processes. Then, it is not relevant to compare joint distributions over a set of electrodes with marginal distributions over individual electrodes.

In the context of the present SMNI study, the CMI transform covariantly under Riemannian transformations, but are more sensitive measures of neocortical activity than other invariants such as the energy density, effectively the square of the CMI, or the information which also effectively is in terms of the square of the CMI (essentially path integrals over quantities proportional to the energy times a factor of an exponential including the energy as an argument). Neither the energy or the information give details of the components as do the CMI. EEG is measuring a quite oscillatory system and the relative signs of such activity are quite important. The information and energy densities are calculated and printed out after ASA fits along with the CMI.

## 4. SMNI APPLICATIONS TO INDIVIDUAL EEG

### 4.1. Data

EEG spontaneous and evoked potential (EP) data from a multi-electrode array under a variety of conditions was collected at several centers in the United States, sponsored by the National Institute on Alcohol Abuse and Alcoholism (NIAAA) project. The earlier 1991 study used only averaged EP data [87]. These experiments, performed on carefully selected sets of subjects, suggest a genetic predisposition to alcoholism that is strongly correlated to EEG AEP responses to patterned targets.

It is clear that the author is not an expert in the clinical aspects of these alcoholism studies. It suffices for this study that the data used is clean raw EEG data, and that these SMNI, CMI, and ASA techniques can and should be used and tested on other sources of EEG data as well.

Each set of results is presented with 6 figures, labeled as [{alcoholic | control}, {stimulus 1 | match | no-match}, subject, {potential | momenta}], abbreviated to {a | c}_{1 | m | n}_subject.{pot | mom} where match or no-match was performed for stimulus 2 after 3.2 sec of a presentation of stimulus 1 [87]. Data includes 10 trials of 69 epochs each between 150 and 400 msec after presentation. For each subject run,

after fitting 28 parameters with ASA, epoch by epoch averages are developed of the raw data and of the multivariate SMNI CMI. It was noted that much poorer fits were achieved when the "centering" mechanism [4,6], driving multiple attractors into the physical firing regions bounded by $M^G \leq \pm N^G$, was turned off and the denominators in $F^G$ were set to constants, confirming the importance of using the full SMNI model. All stimuli were presented for 300 msec. For example, c_m_co2c0000337.pot is a figure. Note that the subject number also includes the {alcoholic | control} tag, but this tag was added just to aid sorting of files (as there are contribution from co2 and co3 subjects). Each figure contains graphs superimposed for 6 electrode sites (out of 64 in the data) which have been modeled by SMNI using a circuitry given in Table 2 of frontal sites (F3 and F4) feeding temporal (sides of head T7 and T8) and parietal (top of head P7 and P8) sites, where odd-numbered (even-numbered) sites refer to the left (right) brain.

---

Table 2

---

### 4.2. ASA Tuning

A three-stage optimization was performed for each of 60 data sets in {a_n, a_m, a_n, c_1, c_m, c_n} of 10 subjects. As described previously, each of these data sets had 3-5 parameters for each SMNI electrode-site model in {F3, F4, T7, T8, P7, P8}, i.e., 28 parameters for each of the optimization runs, to be fit to over 400 pieces of potential data.

For each state generated in the fit, prior to calculating the Lagrangian, tests were performed to ensure that all short-ranged and long-ranged firings lay in their physical boundaries. When this test failed, the generated state was simply excluded from the parameter space for further consideration. This is a standard simulated-annealing technique to handle complex constraints.

#### 4.2.1. First-Stage Optimization

The first-stage optimization used ASA, version 13.1, tuned to give reasonable performance by examining intermediate results of several sample runs in detail. Table 3 gives those OPTIONS changed from their defaults. (See Appendix A for a discussion of ASA OPTIONS.)

________________________________________________________________________

Table 3

________________________________________________________________________

The ranges of the parameters were decided as follows. The ranges of the strength of the long-range connectivities $d_v$ were from 0 to 1. The ranges of the $\{a, b, \phi\}$ parameters were decided by using minimum and maximum values of $M^G$ and $M^{\ddagger G}$ firings to keep the potential variable within the minimum and maximum values of the experimentally measured potential at each electrode site. (This corrects a typo in a previous paper [21], where these 3 parameters were referred to as $\{a, b, c\}$.)

Using the above ASA OPTIONS and ranges of parameters, it was found that typically within several thousand generated states, the global minimum was approached within at least one or two significant figures of the effective Lagrangian (including the prefactor). This estimate was based on final fits achieved with hundreds of thousands of generated states. Runs were permitted to continue for 50,000 generated states. This very rapid convergence in these 30-dimensional spaces was partially due to the invocation of the centering mechanism.

Some tests with SMNI parameters off the diagonal in $M^G$-space, as established by the centering mechanism, confirmed that ASA converged back to this diagonal, but requiring many more generated states. Of course, an examination of the Lagrangian shows this trivially, as noted in previous papers [3,4], wherein the Lagrangian values were on the order of $10^5 \, \tau^{-1}$, compared to $10^{-2}$–$10^{-3} \, \tau^{-1}$ along the diagonal established by the centering mechanism.

### 4.2.2. Second-Stage Optimization

The second-stage optimization was invoked to minimize the number of generated states that would have been required if only the first-stage optimization were performed. Table 4 gives the changes made in the OPTIONS from stage one for stage two. The final stage-one parameters were used as the initial starting parameters for stage two. (At high annealing/quenching temperatures at the start of an SA run, it typically is not important as to what the initial values of the the parameters are, provided of course that they satisfy all constraints, etc.) The second-stage minimum of each parameter was chosen to be the maximum lower bound of the first-stage minimum and a 20% increase of that minimum. The second-stage maximum of each parameter was chosen to be the minimum upper bound of the first-stage maximum and a 20% decrease of that maximum.

___

Table 4

___

Extreme quenching was turned on for the parameters (not for the cost temperature), at values of the parameter dimension of 30, increased from 1 (for rigorous annealing). This worked very well, typically achieving the global minimum with 1000 generated states. Runs were permitted to continue for 10000 generated states.

### 4.2.3. Third-Stage Optimization

The third-stage optimization used a quasi-local code, the Broyden-Fletcher-Goldfarb-Shanno (BFGS) algorithm [88], to gain an extra 2 or 3 figures of precision in the global minimum. This typically took several hundred states, and runs were permitted to continue for 500 generated states. Constraints were enforced by the method of penalties added to the cost function outside the constraints.

The BFGS code typically got stuck in a local minimum quite early if invoked just after the first-stage optimization. (There never was a reasonable chance of getting close to the global minimum using the BFGS code as a first-stage optimizer.) These fits were much more efficient than those in a previous 1991 study [13], where VFSR, the precursor code to ASA, was used for a long stage-one optimization which was then turned over to BFGS.

Table 5 gives the 28 ASA-fitted parameters for each of the 3 experimental paradigms for each alcoholic and control subject.

___

Table 5

___

### 4.3. Testing Data

When the parameters of a theory of a physical system posses clear relationships to observed physical entities, and the theory fits experimental phenomenon while the parameters stay within experimentally determined ranges of these entities, then generally it is conceded that the theory and its parameters have passed a reasonable test. It is argued that this is the case for SMNI and its parameters, and this approach sufficed for the first study of the present data [21], just as SMNI also has been tested in previous papers.

When a model of a physical system has a relatively phenomenological nature then often such a model is best tested by first "training" its parameters on one set of data, then seeing to what degree the same parameters can be used to match the model to out-of-sample "testing" data. For example, this was performed for the statistical mechanics of financial markets (SMFM) project [70-72], applied to trading models [22,23]. The SMFM projects similarly use ASA and the algebra presented here for this SMNI project.

In the present project, there exists barely enough data to additionally test SMNI in this training versus testing methodology. That is, when first examining the data, it was decided to to try to find sets of data from at least 10 control and 10 alcoholic subjects, each set containing at least 10 runs for each of the 3 experimental paradigms, as reported in a previous paper [21]. When reviewing this data, e.g., for the example of the one alcoholic and the one control subject which were illustrated in graphs in that previous paper, it was determined that there exists 10 additional sets of data for each subject for each paradigm, except for the c_n case of the no-match paradigm for the control subject where only 5 additional out-of-sample runs exist. For this latter case, to keep the number of runs sampled consistent across all sets of data, e.g., to keep the relative amplitudes of fluctuations reasonably meaningful, 5 runs of the previous testing set were joined with the 5 runs of the present training set to fill out the data sets required for this study.

### 4.4. Results

Figs. 1-3 compares the CMI to raw data for an alcoholic subject for the a_1, a_m and a_n paradigms, for both the training and testing data. Figs. 4-6 gives similar comparisons for a control subject for the c_1, c_m and c_n paradigms. The SMNI CMI clearly give better signal to noise resolution than the raw data, especially comparing the significant matching tasks between the control and the alcoholic groups, e.g., the c_m and a_m paradigms, in both the training and testing cases. The CMI can be processed further as is the raw data, and also used to calculate "energy" and "information/entropy" densities.

___

Figures 1-6

___

Similar results are seen for other subjects. A compressed tarfile of additonal results for 10 control and 10 alcoholic subjects, using the training data and the testing data, including tables of ASA-fitted parameters and 60 files containing 240 PostScript graphs, can be retrieved via WWW from http://www.ingber.com/MISC.DIR/smni97_eeg_cmi.tar.Z, or as file smni96_eeg_cmi.tar.Z via FTP from ftp.ingber.com in the MISC.DIR directory.

After the above training-testing methodology is applied to more subjects, it will then be possible to perform additional statistical analyses to seek more abbreviated measures of differences between alcoholic and control groups across the 3 experimental paradigms.

**4.5. Availability of Codes and Data**

The ASA code can be downloaded at no charge from http://www.ingber.com/ under WWW or from ftp://ftp.ingber.com under FTP. A mirror site for the home page is http://www.alumni.caltech.edu/˜ingber/ under WWW.

This ASA applications to EEG analysis is one of several ASA applications being prepared for the SPEC (Standard Performance Evaluation Corporation) CPU98 suite. The goal of the program is to elicit benchmarks representing important applications in various technical fields. When SPEC publishes its CDROM, the SMNI CMI codes will be included.

It is extremely difficult for modelers of nonlinear time series, and EEG systems in particular, to get access to large sets of raw clean data. The data used for this study is now publicly available, as described in http://www.ingber.com/smni_eeg_data.html under WWW or ftp://ftp.ingber.com/MISC.DIR/smni_eeg_data.txt under FTP. Data is being provided as is, and may be deleted without notice, or moved to other archives with the only notice given in that file. There are 11,075 typical gzip'd files in the 122 tar'd directories; each file is about 65K. The entire set of data is about 700 MBytes in this compressed format. For convenience, there are subsets of data available that were used for this study.

## 5.  SOURCE LOCALIZATION

### 5.1.  Global and Local Dynamics

The SMNI view is that, for purposes of regional EEG activity, short-ranged interactions can be treated statistically, while long-ranged interactions can be incorporated more specifically by fitting explicit circuitries.  A different approach is taken by present global approaches, which include some statistics of long-range circuitries to develop a wave theory over all or most of neocortex [38,40,89].  The SMNI approach can provide specific fitted long-ranged circuits as points to insert into such global approaches, e.g., to use as specific "data" points to fit the global waves.

For example, a leading global theory [38,40] consists of three coupled equations, two linear and one nonlinear.  The first two describe the number of synaptic events in a volume as being linearly related to the number of action potentials in the brain volume (with a space-dependent time delay).  These equations can be represented in either (space, time) or (wavenumber, frequency) dimensions.  Most of the complication ultimately comes from the third equation, relating action potentials fired to synaptic input.  This is where a better (nonlinear) description is needed.  SMNI can provide this as it is articulated in terms of Lagrangians and probability densities in real space and real time, e.g., of a field of neuronal columns throughout neocortex.  Interactions among these columns is in terms of local and global columnar firings of afferent convergent and efferent diverging fibers.  Synaptic activities are an intermediary calculation in SMNI.

These complementary global and local approaches can be integrated, e.g.,

$$\text{EEG Data -> local theory -> columnar firings -> global theory}$$

For example, in one approach, Laplacians of EEG potentials over a pre-determined brain surface could be used as variables input into SMNI to fit neuronal-synaptic parameters via the joint action $A$ (the time integral over the Lagrangian $L$) integrated over the brain surface.  These firings would then be transformed via global theory into synaptic activities, from which dispersion relations deduce global dynamics [38,40].

## 5.2. Approaches to Source Localization

As described above, SMNI stresses that is natural to describe regional macroscopic dynamics in terms of mesocolumnar interactions, i.e., a true "field" of firings is developed, which has been mathematically articulated in the SMNI papers as defining the regional propagator in the path integral $D\tilde{M}$ over the space-time volume $dx-dt$ of the brain, in terms of the Lagrangian $L$ each volume point. The space volume $dx$ is where the head shape explicitly enters the calculation. For example, fitting $L$ to actual data might require discretization at electrode sites, as performed in this study, where actually the cost function in terms of $L$ to be fit contains the volume elements in $dx$. If a 3-D fit were being done, then $dx$ would be the head volume; if a 2-D surface fit were being done, $dx$ would be the head/brain surface. In practice, we must rely on other methods, e.g., MRI, to determine the head shape to articulate the shape spanned by $dx$. A higher resolution algorithm, at the level of minicolumns, shows how short-ranged as well as long-ranged interactions among mesocolumns within and between regions can be naturally included in the SMNI theory [14].

Such fits can be performed after other source localization techniques are applied to the data, e.g., head-volume or Laplacian techniques. This still is very important, as identification of source(s), especially those that may be nonstationary, is not sufficient to describe their dynamical interaction. The fitted SMNI dynamics then can provide the dynamical description to predict future evolution of the interactions among sources, e.g., to fill in gaps in data that might aid correlation with behavioral states.

The SMNI dynamics can more directly be part of source localization algorithms. For example, if $M/dt$ is identified as the the mesocolumnar current ($M$ is essentially the number of neuronal firings within the time $\tau$ of about 5 msec; $dt$ is the time resolution of the EEG, typically on this order or somewhat less), then we can identify $M = B\nabla^2\Phi$, where $\nabla^2$ is the Laplacian and $B$ is a constant included in the fit, similar to the relationship between $M$ and $\Phi$ in this present project. Then, the Laplacian of $\Phi$ would be input into the SMNI action $A$, and the parameters of the model would be fit as performed here. The ASA global optimization of the highly nonlinear SMNI finds the best fit among the combined local-global interactions algebraically described in the SMNI action $A$. The "curvatures" (second derivatives) of the parameters about the global minimum, automatically returned by ASA, give a covariance matrix of the goodness of fit about the global minimum.

To include some global dynamics in the language of present global theories, e.g., if some subset of Legendre-decomposed $\Phi$ are deemed to be most important, their coefficients can be considered parameters, and the fit can find the degree of which partial waves are most likely present in the data.

It is important to stress that the above approach recognizes that both local and global neuronal dynamics, head shapes, etc., enter as nonlinear stochastic events in the real brain, and this approach can match these events, but it requires global optimization to get the specificity to fit real individual brains.

## 6. CONCLUSIONS

### 6.1. CMI and Linear Models

It is clear that the CMI follow the measured potential variables closely. In large part, this is due to the prominent attractors near the firing states $M^G$ being close to their origins, resulting in moderate threshold functions $F^G$ in these regions. This keeps the term in the drifts proportional to $\tanh F^G$ near its lowest values, yielding values of the drifts on the order of the time derivatives of the potentials. The diffusions, proportional to $\text{sech} F^G$, also do not fluctuate to very large values.

However, when the feedback among potentials under electrode sites are strong, leading to enhanced (nonlinear) changes in the drifts and diffusions, then these do cause relatively largest signals in the CMI relative to those appearing in the raw potentials. Thus, these effects are strongest in the c_m sets of data, where the control (normal) subjects demonstrate more intense circuitry interactions among electrode sites during the matching paradigm.

These results also support independent studies of primarily long-ranged EEG activity, that have concluded that EEG many times appears to demonstrate quasi-linear interactions [40,90]. However, it must be noted that this is only true within the confines of an attractor of highly nonlinear short-ranged columnar interactions. It requires some effort, e.g., global optimization of a robust multivariate stochastic nonlinear system to achieve finding this attractor. Theoretically, using the SMNI model, this is performed using the ASA code. Presumably, the neocortical system utilizes neuromodular controls to achieve this attractor state [56,81], as suggested in early SMNI studies [3,4].

## 6.2. CMI Features

Essential features of the SMNI CMI approach are: (a) A realistic SMNI model, clearly capable of modeling EEG phenomena, is used, including both long-ranged columnar interactions across electrode sites and short-ranged columnar interactions under each electrode site. (b) The data is used raw for the nonlinear model, and only after the fits are moments (averages and variances) taken of the derived CMI indicators; this is unlike other studies that most often start with averaged potential data. (c) A novel and sensitive measure, CMI, is used, which has been shown to be successful in enhancing resolution of signals in another stochastic multivariate time series system, financial markets [22,23]. As was performed in those studies, future SMNI projects can similarly use recursive ASA optimization, with an inner-shell fitting CMI of subjects' EEG, embedded in an outer-shell of parameterized customized clinician's AI-type rules acting on the CMI, to create supplemental decision aids.

Canonical momenta offers an intuitive yet detailed coordinate system of some complex systems amenable to modeling by methods of nonlinear nonequilibrium multivariate statistical mechanics. These can be used as reasonable indicators of new and/or strong trends of behavior, upon which reasonable decisions and actions can be based, and therefore can be be considered as important supplemental aids to other clinical indicators.

## 6.3. CMI and Source Localization

Global ASA optimization, fitting the nonlinearities inherent in the synergistic contributions from short-ranged columnar firings and from long-ranged fibers, makes it possible to disentangle their contributions to some specific electrode circuitries among columnar firings under regions separated by cm, at least to the degree that the CMI clearly offer superior signal to noise than the raw data. Thus this paper at least establishes the utility of the CMI for EEG analyses, which can be used to complement other EEG modeling techniques. In this paper, a plausible circuitry was first hypothesized (by a group of experts), and it remains to be seen just how many more electrodes can be added to such studies with the goal being to have ASA fits determine the optimal circuitry.

### 6.4. SMNI Features

Sets of EEG data taken during selective attention tasks have been fit using parameters either set to experimentally observed values, or have been fit within experimentally observed values. The ranges of columnar firings are consistent with a centering mechanism derived for STM in earlier papers.

These results, in addition to their importance in reasonably modeling EEG with SMNI, also have a deeper theoretical importance with respect to the scaling of neocortical mechanisms of interaction across disparate spatial scales and behavioral phenomena: As has been pointed out previously, SMNI has given experimental support to the derivation of the mesoscopic probability distribution, illustrating common forms of interactions between their entities, i.e., neurons and columns of neurons, respectively. The nonlinear threshold factors are defined in terms of electrical-chemical synaptic and neuronal parameters all lying within their experimentally observed ranges. It also was noted that the most likely trajectories of the mesoscopic probability distribution, representing averages over columnar domains, give a description of the systematics of macroscopic EEG in accordance with experimental observations. It has been demonstrated that the macroscopic regional probability distribution can be derived to have same functional form as the mesoscopic distribution, where the macroscopic drifts and diffusions of the potentials described by the $\Phi$'s are simply linearly related to the (nonlinear) mesoscopic drifts and diffusions of the columnar firing states given by the $M^G$'s. Then, this macroscopic probability distribution gives a reasonable description of experimentally observed EEG.

The theoretical and experimental importance of specific scaling of interactions in the neocortex has been quantitatively demonstrated on individual brains. The explicit algebraic form of the probability distribution for mesoscopic columnar interactions is driven by a nonlinear threshold factor of the same form taken to describe microscopic neuronal interactions, in terms of electrical-chemical synaptic and neuronal parameters all lying within their experimentally observed ranges; these threshold factors largely determine the nature of the drifts and diffusions of the system. This mesoscopic probability distribution has successfully described STM phenomena and, when used as a basis to derive the most likely trajectories using the Euler-Lagrange variational equations, it also has described the systematics of EEG phenomena. In this paper, the mesoscopic form of the full probability distribution has been taken more seriously for macroscopic interactions, deriving macroscopic drifts and diffusions linearly related to sums of their (nonlinear) mesoscopic counterparts, scaling its variables to describe interactions among regional interactions correlated with observed electrical activities measured by electrode recordings of scalp EEG, with apparent success. These results give strong quantitative support for an accurate intuitive picture, portraying neocortical interactions as having common algebraic or physics mechanisms that scale across quite disparate spatial scales and functional or behavioral phenomena, i.e., describing interactions among neurons, columns of neurons, and regional masses of neurons.

**6.5. Summary**

SMNI is a reasonable approach to extract more ''signal'' out of the ''noise'' in EEG data, in terms of physical dynamical variables, than by merely performing regression statistical analyses on collateral variables. To learn more about complex systems, inevitably functional models must be formed to represent huge sets of data. Indeed, modeling phenomena is as much a cornerstone of 20th century science as is collection of empirical data [91].

It seems reasonable to speculate on the evolutionary desirability of developing Gaussian-Markovian statistics at the mesoscopic columnar scale from microscopic neuronal interactions, and maintaining this type of system up to the macroscopic regional scale. I.e., this permits maximal processing of information [74]. There is much work to be done, but modern methods of statistical mechanics have helped to point the way to promising approaches.

**APPENDIX A: ADAPTIVE SIMULATED ANNEALING (ASA)**

**1. General Description**

Simulated annealing (SA) was developed in 1983 to deal with highly nonlinear problems [92], as an extension of a Monte-Carlo importance-sampling technique developed in 1953 for chemical physics problems. It helps to visualize the problems presented by such complex systems as a geographical terrain. For example, consider a mountain range, with two "parameters," e.g., along the North–South and East–West directions, with the goal to find the lowest valley in this terrain. SA approaches this problem similar to using a bouncing ball that can bounce over mountains from valley to valley. Start at a high "temperature," where the temperature is an SA parameter that mimics the effect of a fast moving particle in a hot object like a hot molten metal, thereby permitting the ball to make very high bounces and being able to bounce over any mountain to access any valley, given enough bounces. As the temperature is made relatively colder, the ball cannot bounce so high, and it also can settle to become trapped in relatively smaller ranges of valleys.

Imagine that a mountain range is aptly described by a "cost function." Define probability distributions of the two directional parameters, called generating distributions since they generate possible valleys or states to explore. Define another distribution, called the acceptance distribution, which depends on the difference of cost functions of the present generated valley to be explored and the last saved lowest valley. The acceptance distribution decides probabilistically whether to stay in a new lower valley or to bounce out of it. All the generating and acceptance distributions depend on temperatures.

In 1984 [93], it was established that SA possessed a proof that, by carefully controlling the rates of cooling of temperatures, it could statistically find the best minimum, e.g., the lowest valley of our example above. This was good news for people trying to solve hard problems which could not be solved by other algorithms. The bad news was that the guarantee was only good if they were willing to run SA forever. In 1987, a method of fast annealing (FA) was developed [94], which permitted lowering the temperature exponentially faster, thereby statistically guaranteeing that the minimum could be found in some finite time. However, that time still could be quite long. Shortly thereafter, Very Fast Simulated Reannealing (VFSR) was developed [25], now called Adaptive Simulated Annealing (ASA), which is exponentially faster than FA.

ASA has been applied to many problems by many people in many disciplines [27,28,95]. The feedback of many users regularly scrutinizing the source code ensures its soundness as it becomes more flexible and powerful. The code is available via the world-wide web (WWW) as http://www.ingber.com/ which also can be accessed anonymous FTP from ftp.ingber.com.

## 2. Mathematical Outline

ASA considers a parameter $\alpha_k^i$ in dimension $i$ generated at annealing-time $k$ with the range

$$\alpha_k^i \in [A_i, B_i] \,, \tag{A.1}$$

calculated with the random variable $y^i$,

$$\alpha_{k+1}^i = \alpha_k^i + y^i(B_i - A_i) \,,$$

$$y^i \in [-1, 1] \ . \tag{A.2}$$

The generating function $g_T(y)$ is defined,

$$g_T(y) = \prod_{i=1}^{D} \frac{1}{2(|y^i| + T_i) \ln(1 + 1/T_i)} \equiv \prod_{i=1}^{D} g_T^i(y^i) \ , \tag{A.3}$$

where the subscript $i$ on $T_i$ specifies the parameter index, and the $k$-dependence in $T_i(k)$ for the annealing schedule has been dropped for brevity. Its cumulative probability distribution is

$$G_T(y) = \int_{-1}^{y^1} \cdots \int_{-1}^{y^D} dy'^1 \cdots dy'^D \, g_T(y') \equiv \prod_{i=1}^{D} G_T^i(y^i) \ ,$$

$$G_T^i(y^i) = \frac{1}{2} + \frac{\operatorname{sgn}(y^i)}{2} \frac{\ln(1 + |y^i|/T_i)}{\ln(1 + 1/T_i)} \ . \tag{A.4}$$

$y^i$ is generated from a $u^i$ from the uniform distribution

$$u^i \in U[0, 1] \ ,$$

$$y^i = \operatorname{sgn}(u^i - \frac{1}{2}) T_i [(1 + 1/T_i)^{|2u^i - 1|} - 1] \ . \tag{A.5}$$

It is straightforward to calculate that for an annealing schedule for $T_i$

$$T_i(k) = T_{0i} \exp(-c_i k^{1/D}) \ , \tag{A.6}$$

a global minima statistically can be obtained. I.e.,

$$\sum_{k_0}^{\infty} g_k \approx \sum_{k_0}^{\infty} [\prod_{i=1}^{D} \frac{1}{2|y^i|c_i}] \frac{1}{k} = \infty \ . \tag{A.7}$$

Control can be taken over $c_i$, such that

$$T_{fi} = T_{0i} \exp(-m_i) \text{ when } k_f = \exp n_i \ ,$$

$$c_i = m_i \exp(-n_i/D) \ , \tag{A.8}$$

where $m_i$ and $n_i$ can be considered "free" parameters to help tune ASA for specific problems.

## 3. ASA OPTIONS

ASA has over 100 OPTIONS available for tuning. A few are most relevant to this project.

### 3.1. Reannealing

Whenever doing a multi-dimensional search in the course of a complex nonlinear physical problem, inevitably one must deal with different changing sensitivities of the $\alpha^i$ in the search. At any given annealing-time, the range over which the relatively insensitive parameters are being searched can be "stretched out" relative to the ranges of the more sensitive parameters. This can be accomplished by periodically rescaling the annealing-time $k$, essentially reannealing, every hundred or so acceptance-events (or at some user-defined modulus of the number of accepted or generated states), in terms of the sensitivities $s_i$ calculated at the most current minimum value of the cost function, $C$,

$$s_i = \partial C/\partial \alpha^i \ . \tag{A.9}$$

In terms of the largest $s_i = s_{\max}$, a default rescaling is performed for each $k_i$ of each parameter dimension, whereby a new index $k'_i$ is calculated from each $k_i$,

$$k_i \to k'_i \ ,$$

$$T'_{ik'} = T_{ik}(s_{\max}/s_i) \ ,$$

$$k'_i = (\ln(T_{i0}/T_{ik'})/c_i)^D \ . \tag{A.10}$$

$T_{i0}$ is set to unity to begin the search, which is ample to span each parameter dimension.

### 3.2. Quenching

Another adaptive feature of ASA is its ability to perform quenching in a methodical fashion. This is applied by noting that the temperature schedule above can be redefined as

$$T_i(k_i) = T_{0i} \exp(-c_i k_i^{Q_i/D}) \ ,$$

$$c_i = m_i \exp(-n_i Q_i/D) \ , \tag{A.11}$$

in terms of the "quenching factor" $Q_i$. The sampling proof fails if $Q_i > 1$ as

$$\sum_k \prod_k^D 1/k^{Q_i/D} = \sum_k 1/k^{Q_i} < \infty \ . \tag{A.12}$$

This simple calculation shows how the "curse of dimensionality" arises, and also gives a possible way of living with this disease. In ASA, the influence of large dimensions becomes clearly focussed on the exponential of the power of $k$ being $1/D$, as the annealing required to properly sample the space becomes prohibitively slow. So, if resources cannot be committed to properly sample the space, then for some systems perhaps the next best procedure may be to turn on quenching, whereby $Q_i$ can become on the order of the size of number of dimensions.

The scale of the power of $1/D$ temperature schedule used for the acceptance function can be altered in a similar fashion. However, this does not affect the annealing proof of ASA, and so this may used without damaging the sampling property.

### 3.3. Self Optimization

If not much information is known about a particular system, if the ASA defaults do not seem to work very well, and if after a bit of experimentation it still is not clear how to select values for some of the ASA OPTIONS, then the SELF_OPTIMIZE OPTIONS can be very useful. This sets up a top level search on the ASA OPTIONS themselves, using criteria of the system as its own cost function, e.g., the best attained optimal value of the system's cost function (the cost function for the actual problem to be solved) for each given set of top level OPTIONS, or the number of generated states required to reach a given value of the system's cost function, etc. Since this can consume a lot of CPU resources, it is recommended that only a few ASA OPTIONS and a scaled down system cost function or system data be selected for this OPTIONS.

Even if good results are being attained by ASA, SELF_OPTIMIZE can be used to find a more efficient set of ASA OPTIONS. Self optimization of such parameters can be very useful for production runs of complex systems.

### 3.4. Parallel Code

It is quite difficult to directly parallelize an SA algorithm [27], e.g., without incurring very restrictive constraints on temperature schedules [96], or violating an associated sampling proof [97]. However, the fat tail of ASA permits parallelization of developing generated states prior to subjecting them to the acceptance test [14]. The ASA_PARALLEL OPTIONS provide parameters to easily parallelize the code, using various implementations, e.g., PVM, shared memory, etc.

The scale of parallelization afforded by ASA, without violating its sampling proof, is given by a typical ratio of the number of generated to accepted states. Several experts in parallelization suggest that massive parallelization e.g., on the order of the human brain, may take place quite far into the future, that this might be somewhat less useful for many applications than previously thought, and that most useful scales of parallelization might be on scales of order 10 to 1000. Depending on the specific problem, such scales are common in ASA optimization, and the ASA code can implement such parallelization.

**ACKNOWLEDGMENTS**

I thank Paul Nunez at Tulane University for several discussions on issues of source localization. Data was collected by Henri Begleiter and associates at the Neurodynamics Laboratory at the State University of New York Health Center at Brooklyn, and prepared by David Chorlian. Calculations were performed on a Sun SPARC 20 at the University of Oregon, Eugene, courtesy of the Department of Psychology, consuming about 200 CPU-hours, and they have made a GByte of disk space available to archive the data from the Neurodynamics Laboratory.

Statistical Mechanics of Neocortical ...          - 41 -                            Lester Ingber

**FIGURE CAPTIONS**

Figure 1. For the initial-stimulus a_1 paradigm for alcoholic subject co2a0000364, each figure gives data under 6 electrodes marked in the legends. The left hand figures represent data for the training calculations; the right hand figures represent data for the testing calculations. The top figures represent averages over 10 runs of raw evoked potential data; the bottom figures represent averages over 10 calculations of canonical momenta indicators using this data.

Figure 2. For the match second-stimulus a_m paradigm for alcoholic subject co2a0000364, each figure gives data under 6 electrodes marked in the legends. Descriptions of the four figures are contained in the legend for Figure 1.

Figure 3. For the no-match second-stimulus a_n paradigm for alcoholic subject co2a0000364, each figure gives data under 6 electrodes marked in the legends. Descriptions of the four figures are contained in the legend for Figure 1.

Figure 4. For the initial-stimulus c_1 paradigm for control subject co2c0000337, each figure gives data under 6 electrodes marked in the legends. Descriptions of the four figures are contained in the legend for Figure 1.

Figure 5. For the match second-stimulus c_m paradigm for control subject co2c0000337, each figure gives data under 6 electrodes marked in the legends. Descriptions of the four figures are contained in the legend for Figure 1.

Figure 6. For the no-match second-stimulus c_n paradigm for control subject co2c0000337, each figure gives data under 6 electrodes marked in the legends. Descriptions of the four figures are contained in the legend for Figure 1.

**TABLE CAPTIONS**

Table 1. Descriptive concepts and their mathematical equivalents in a Lagrangian representation. These physical entities provide another form of intuitive, but quantitatively precise, presentation of these analyses.

Table 2. Circuitry of long-ranged fibers across most relevant electrode sites and their assumed time-delays in units of 3.906 msec.

Table 3. ASA OPTIONS changes from their defaults used in stage one optimization.

Table 4. ASA OPTIONS changes from their use in stage one for stage two optimization.

Table 5. Parameters fit by ASA are given, as described in the text, for 3 sets of data for alcoholic subject co2a0000364 and for control subject co2c0000337.

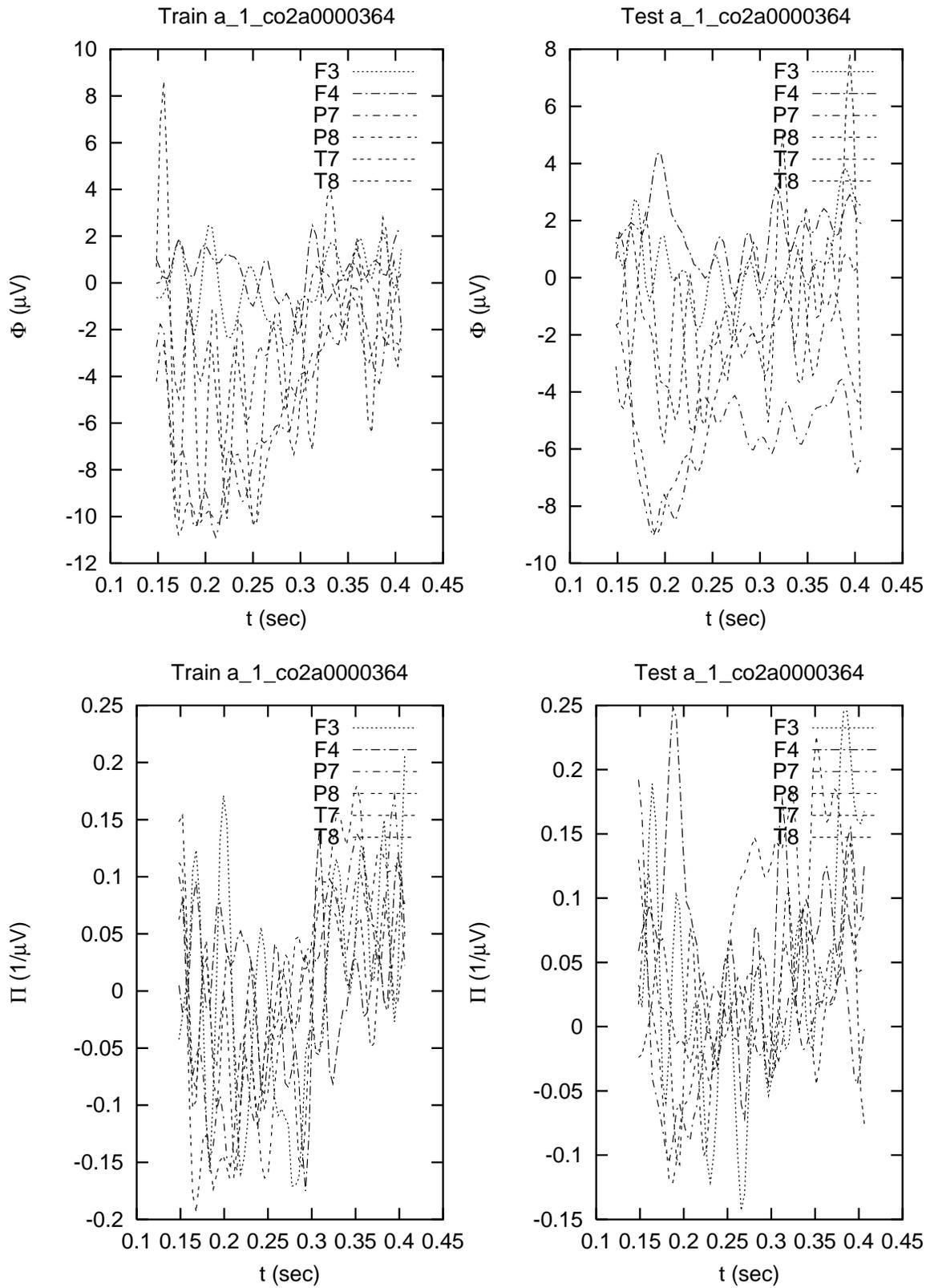

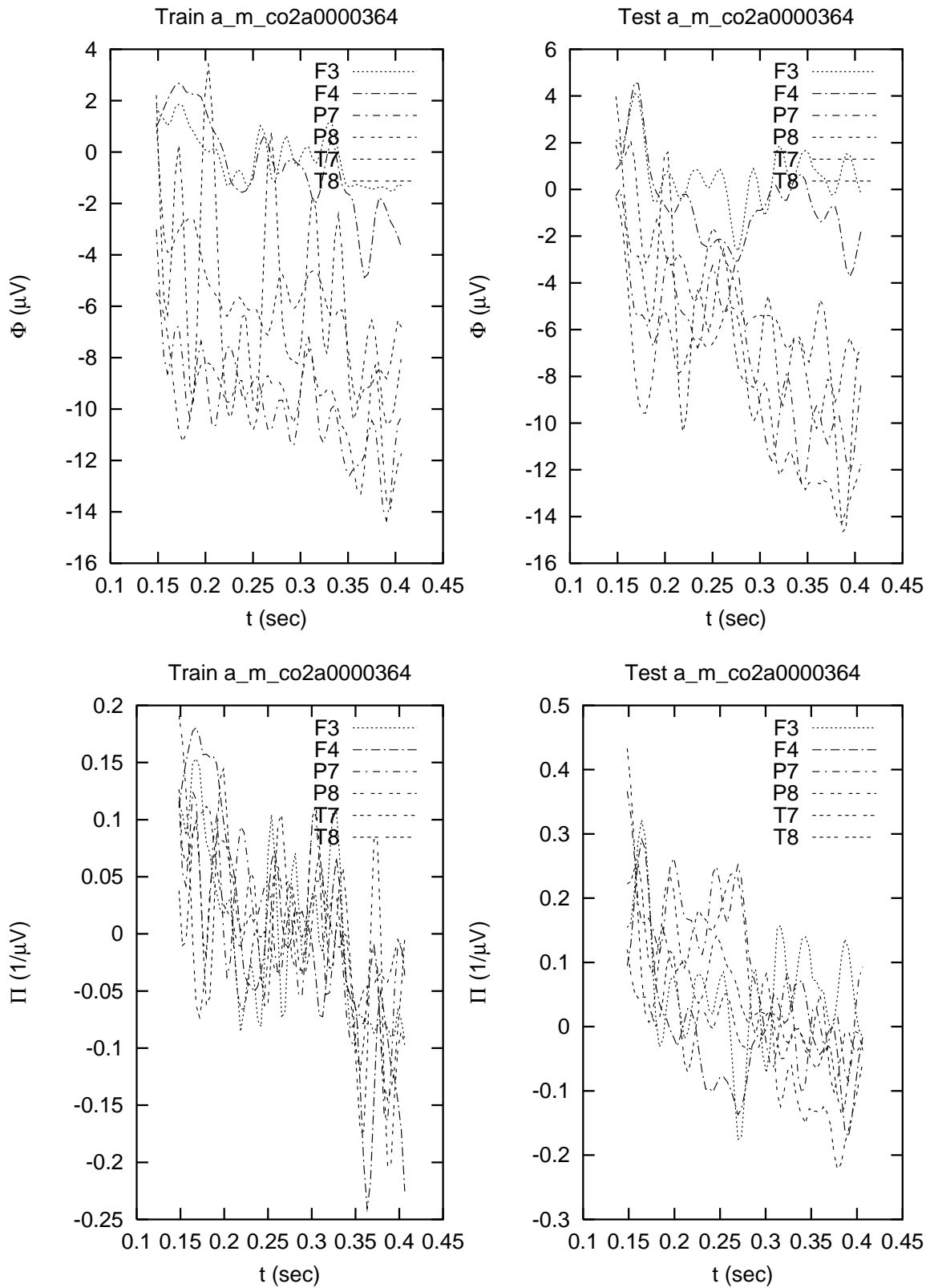

Statistical Mechanics of Neocortical ...    - Figure 3 -    Lester Ingber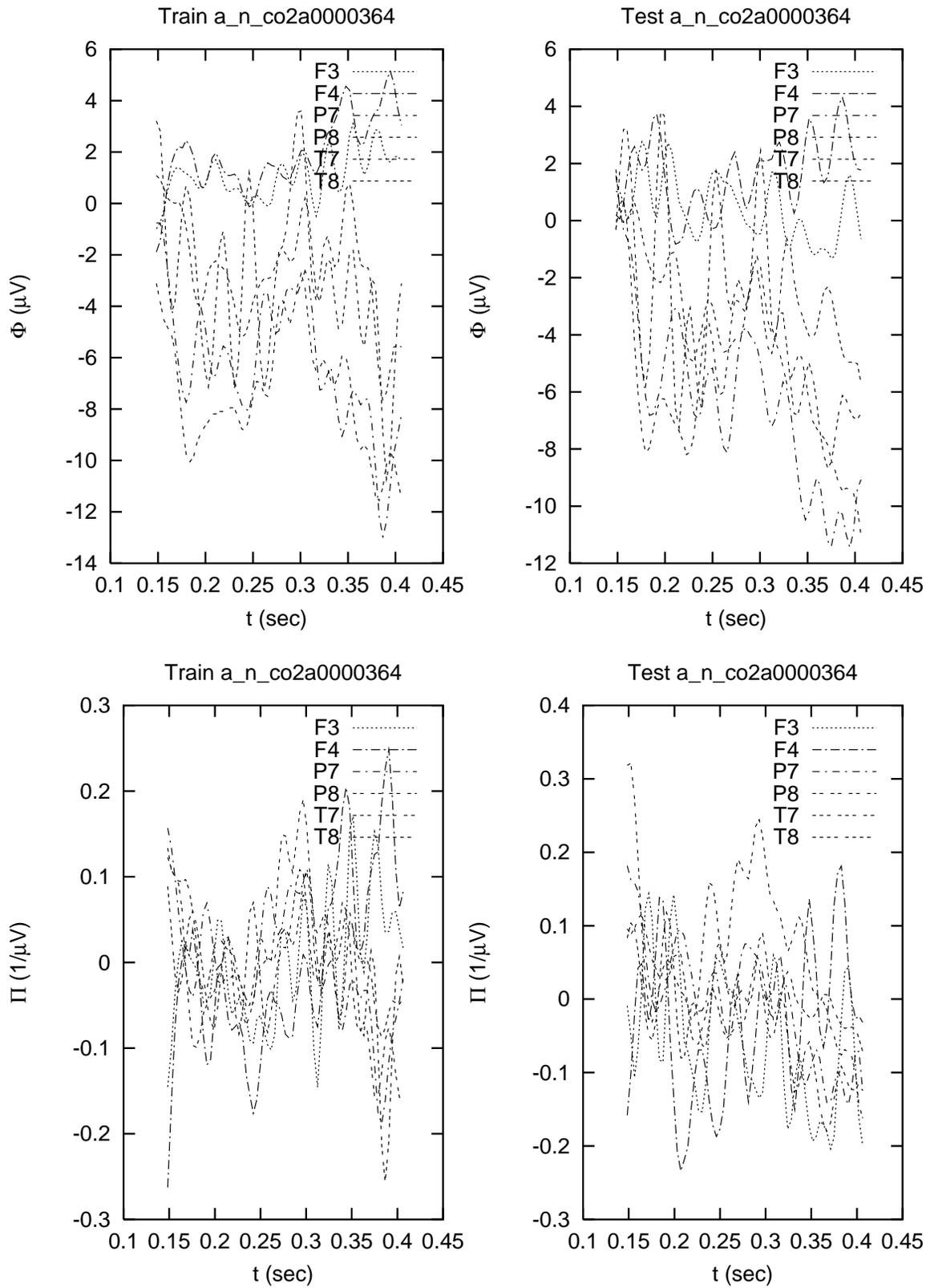

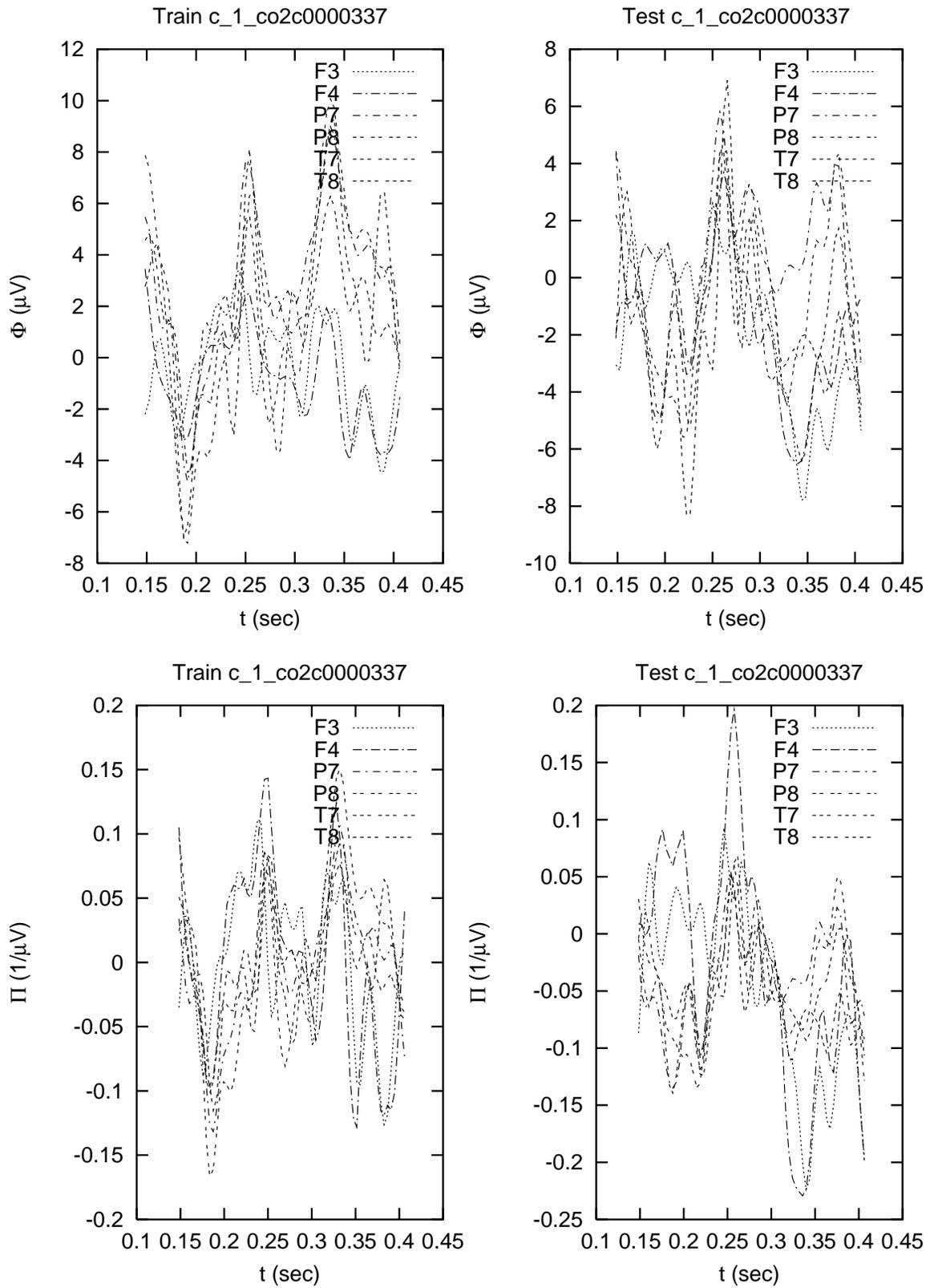

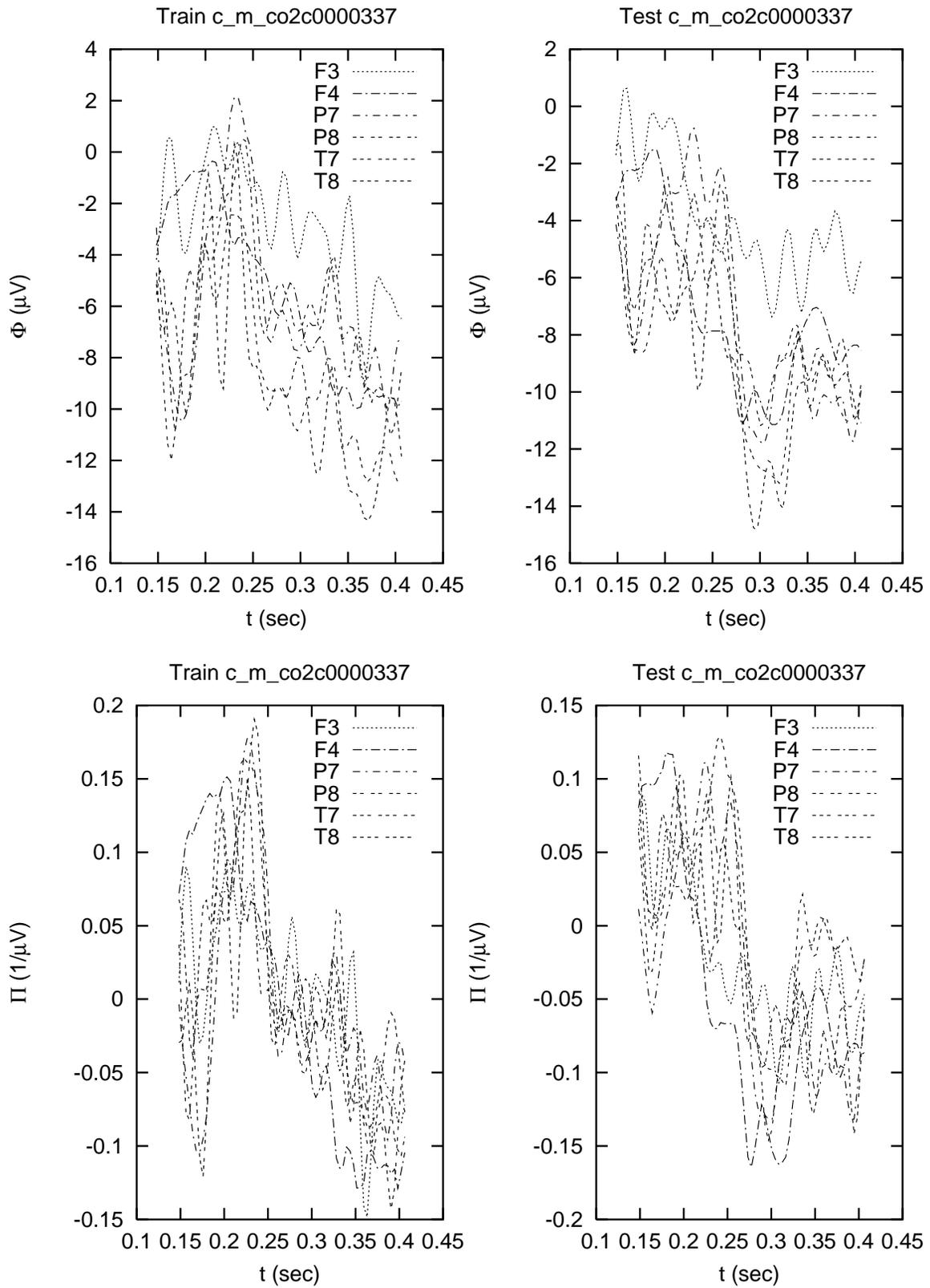

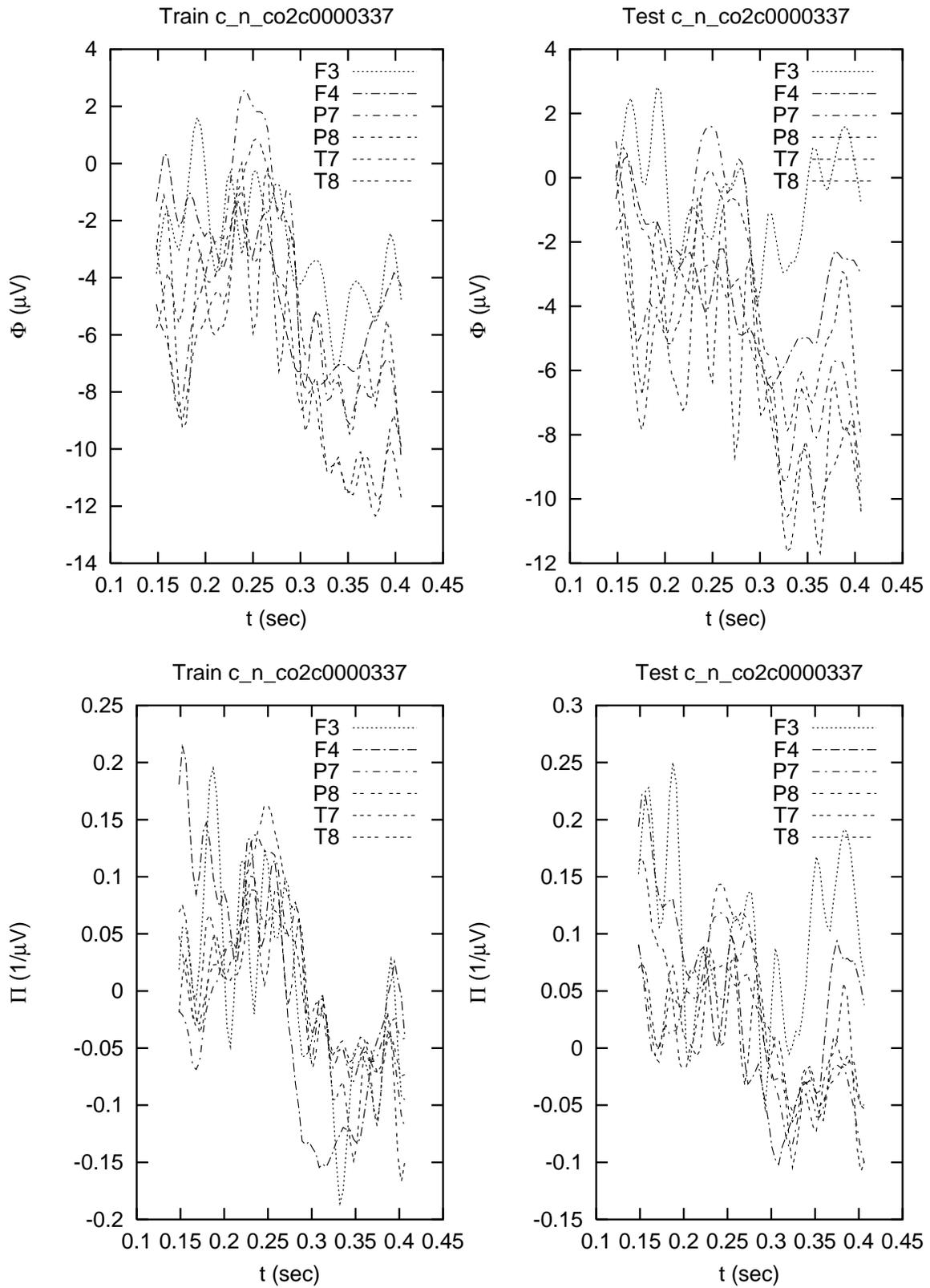

| Concept | Lagrangian equivalent |
|---|---|
| Momentum | $\Pi^G = \dfrac{\partial \underline{L}_F}{\partial(\partial M^G/\partial t)}$ |
| Mass | $g_{GG'} = \dfrac{\partial \underline{L}_F}{\partial(\partial M^G/\partial t)\partial(\partial M^{G'}/\partial t)}$ |
| Force | $\dfrac{\partial \underline{L}_F}{\partial M^G}$ |
| $F = ma$ | $\delta \underline{L}_F = 0 = \dfrac{\partial \underline{L}_F}{\partial M^G} - \dfrac{\partial}{\partial t}\dfrac{\partial \underline{L}_F}{\partial(\partial M^G/\partial t)}$ |

| Site | Contributions From | Time Delays (3.906 msec) |
|------|--------------------|--------------------------|
| F3   | –                  | –                        |
| F4   | –                  | –                        |
| T7   | F3                 | 1                        |
| T7   | T8                 | 1                        |
| T8   | F4                 | 1                        |
| T8   | T7                 | 1                        |
| P7   | T7                 | 1                        |
| P7   | P8                 | 1                        |
| P7   | F3                 | 2                        |
| P8   | T8                 | 1                        |
| P8   | P7                 | 1                        |
| P8   | F4                 | 2                        |

| OPTIONS | Default | Stage 1 Use |
|---|---|---|
| Limit_Acceptances | 10000 | 25000 |
| Limit_Generated | 99999 | 50000 |
| Cost_Precision | 1.0E-18 | 1.0E-9 |
| Number_Cost_Samples | 5 | 3 |
| Cost_Parameter_Scale_Ratio | 1.0 | 0.2 |
| Acceptance_Frequency_Modulus | 100 | 25 |
| Generated_Frequency_Modulus | 10000 | 10 |
| Reanneal_Cost | 1 | 4 |
| Reanneal_Parameters | 1 | 0 |
| SMALL_FLOAT | 1.0E-18 | 1.0E-30 |
| ASA_LIB | FALSE | TRUE |
| QUENCH_COST | FALSE | TRUE |
| QUENCH_PARAMETERS | FALSE | TRUE |
| COST_FILE | TRUE | FALSE |
| NO_PARAM_TEMP_TEST | FALSE | TRUE |
| NO_COST_TEMP_TEST | FALSE | TRUE |
| TIME_CALC | FALSE | TRUE |
| ASA_PRINT_MORE | FALSE | TRUE |

| OPTIONS | Stage 2 Changes |
|---|---|
| Limit_Acceptances | 5000 |
| Limit_Generated | 10000 |
| User_Initial_Parameters | TRUE |
| User_Quench_Param_Scale[.] | 30 |

| Parameter | c_1 | c_n | c_m | a_1 | a_n | a_m |
|---|---|---|---|---|---|---|
| F3 | | | | | | |
| $a$ | -0.281 | 0.174 | -0.346 | 0.245 | -0.342 | 0.139 |
| $b$ | 0.970 | -0.848 | 1.045 | 0.606 | -0.312 | -0.645 |
| $<\phi>$ | -0.297 | -2.930 | -2.922 | -0.015 | 1.127 | -0.260 |
| F4 | | | | | | |
| $a$ | -0.255 | 0.243 | 0.356 | -0.401 | 0.127 | -0.066 |
| $b$ | 0.799 | -0.765 | -0.906 | -0.303 | -0.625 | 0.778 |
| $<\phi>$ | -0.935 | -4.296 | -5.557 | 0.427 | 2.047 | -0.650 |
| T7 | | | | | | |
| $a$ | -0.784 | 0.181 | 0.592 | 0.267 | -0.146 | 0.232 |
| $b$ | -0.552 | -1.035 | 0.500 | -0.973 | 1.005 | -0.961 |
| $<\phi>$ | 1.902 | -4.622 | -6.080 | -2.143 | -3.023 | -6.085 |
| $d_1(F3)$ | 0.325 | 0.993 | 0.579 | 0.306 | 0.898 | 0.589 |
| $d_1(T8)$ | 0.358 | 0.070 | 0.437 | 0.056 | 0.288 | 0.325 |
| T8 | | | | | | |
| $a$ | 0.959 | 0.781 | 0.755 | 0.892 | -0.348 | -0.398 |
| $b$ | 0.624 | 0.989 | 0.785 | 0.701 | 1.295 | 1.355 |
| $<\phi>$ | 1.417 | -6.822 | -8.742 | -2.928 | -2.638 | -6.005 |
| $d_1(F4)$ | 0.441 | 0.101 | 0.468 | 0.814 | 0.138 | 0.255 |
| $d_1(T7)$ | 0.376 | 0.074 | 0.383 | 0.613 | 0.851 | 0.308 |
| P7 | | | | | | |
| $a$ | 0.830 | 0.778 | 0.273 | 0.166 | -0.262 | 0.385 |
| $b$ | 0.371 | 0.498 | -1.146 | -1.083 | 1.006 | 0.720 |
| $<\phi>$ | 2.832 | -4.837 | -5.784 | -5.246 | -6.395 | -9.980 |
| $d_1(T7)$ | 0.291 | 0.257 | 0.153 | 0.157 | 0.472 | 0.081 |
| $d_1(P8)$ | 0.217 | 0.079 | 0.717 | 0.945 | 0.826 | 0.186 |
| $d_2(F3)$ | 0.453 | 0.604 | 0.401 | 0.867 | 0.517 | 0.096 |
| P8 | | | | | | |
| $a$ | 0.350 | 0.300 | 0.618 | 0.155 | -0.534 | 0.130 |
| $b$ | -1.238 | -1.378 | 0.545 | -0.942 | -0.338 | -0.795 |
| $<\phi>$ | 1.772 | -7.231 | -8.866 | -5.641 | -7.539 | -9.869 |
| $d_1(T8)$ | 0.297 | 0.844 | 0.645 | 0.547 | 0.864 | 0.706 |
| $d_1(P7)$ | 0.809 | 0.634 | 0.958 | 0.314 | 0.862 | 0.374 |
| $d_2(F4)$ | 0.569 | 0.600 | 0.487 | 0.426 | 0.282 | 0.722 |